# Current status of the LHCf experiment

O.Adriani[9], L.Bonechi[9], M.Bongi[9], G.Castellini[9], R..D'Alessandro[9], A.Faus[11], K.Fukui[1],
M.Grandi[9], M.Haguenauer[7], Y.Itow[1], K.Kasahara[3], D.Macina[12], T.Mase[1], K.Masuda[1],
Y.Matsubara[1], H.Menjo[1], M.Mizuishi[3], Y.Muraki[5], P.Papini[9], A-L.Perrot[12], S.Ricciarini[9],
T.Sako[1], Y.Shimizu[6], K.Taki[1], T.Tamura[4], S.Torii[3], A.Tricomi[10], W.C.Turner[8],
J.Velasco[11], A. Viciani[9], K.Yoshida[2]    - The LHCf collaboration –

*Abstract*— **An important experiment for cosmic ray physics is going to be conducted with the colliding proton beams of the CERN LHC. The equivalent energy of the 14 TeV center of mass energy of the colliding proton beams in the LHC is $10^{17}$eV in the laboratory frame. Two small electromagnetic calorimeters have been installed at zero degree collision angle +/-140m from the interaction point IP1. The calorimeters measure the energies and the positions of the axes of showers produced by forward neutral particles created in the proton-proton collisions. The shower calorimeters are therefore able to measure the production cross-section of neutral pions emitted in the very forward region. The experimental results will be obtained during the early operation phase of LHC when the luminosity is below $10^{30}$cm$^{-2}$s$^{-1}$. The data obtained will be extremely important for benchmarking the various Monte Carlo codes used for interpreting super high energy cosmic ray phenomena.**

## 1. A BRIEF HISTORY OF NUCLEAR INTERACTION STUDIES BY COSMIC RAY EXPERIMENTS

AROUND 1980, there was a considerable debate in the field of high energy cosmic ray research over whether or not Feynman scaling was violated in the very forward region of collisions above 100 TeV[1]. However another interpretation of the same data was possible; namely that Feynman scaling was not violated, but the composition of primary cosmic rays changed from proton dominated to iron dominated when primary energies exceeded about 100 TeV [2].

Let us show an example presented by the Chacaltaya emulsion chamber group in 1983 [3] in which an apparent violation of Feynman scaling is resolved by a shift of the pseudo-rapidity axis by 20%. Figure 1a represents the pseudo-rapidity distribution of the gamma-rays observed by the Chacaltaya emulsion chamber (the round mark). They are presented in Figure 1a together with CERN ISR experimental data. The expected pseudo-rapidity distribution of the gamma-rays from neutral pions of the CERN ISR experiment is shown on the same graph (the thick curve).

In the Chacaltaya emulsion chamber experiment, a thick carbon target (23cm) was located 150cm above the chamber. High energy protons that penetrated the atmosphere made nuclear interactions in the carbon target producing a number of high energy gamma-rays and other particles. The gamma-rays with energy higher than 200 GeV were detectable by the emulsion chamber. Compared to the ISR data, the Chacaltaya result (open circles and dotted line in Fig. 1a) were shifted in pseudo-rapidity. This was interpreted by the Chacaltaya group as evidence of a violation of Feynman scaling in the very forward region.

However we know that it is difficult to measure the energy of the incident particle accurately with emulsion chambers, since they measure only photons and some fraction of the energy of hadrons. Therefore the pseudo-rapidity distribution could be shifted either toward the right or the left. At least a 20% shift of the horizontal scale in Fig. 1a is possible. If we shift the pseudo-rapidity distribution 20% towards the left side, then the distribution is in accordance with the ISR data as shown in Fig. 1b. In this case, even the Chacaltaya emulsion chamber experiment demonstrates the Feynman scaling at very forward region. It may hold in a wide range of energy from 2 TeV to 100 TeV. To confirm this point we needed to perform an experiment at fixed primary particle energy. Therefore we proposed an accelerator experiment to CERN in 1984 using SPS pbar-p collider in.

## 2. THE CERN UA7 EXPERIMENT

The UA7 experiment was conducted during 1985-1986 with the support of the UA4 collaboration. Two 4 inch silicon calorimeters (ch. A and C) and a 3 inch calorimeter (ch. B) were installed near the intersection point of UA2-UA4 as shown in Figure 2. Chamber A was located about 20m from the intersection point and installed in the position of a Roman pot, while the chambers B and C were located 12m away from the intersection point and outside the beam pipe. With this configuration we measured the photon distribution with the emission angle between 1.8 and 3.4 milli-radians and between 7 and 13 milli-radians from the beam axis respectively.

The results of the UA7 experiment are published in ref. [4]. Some of these results are shown in Figures 3 and 4. Figure 3 represents the invariant mass distribution of two photons detected by the 4" calorimeter that was installed inside the

1) STE laboratory, Nagoya University, Nagoya, 464-801, Japan
2) Shibaura Institute of Technology, Saitama, Japan
3) Physics department, Waseda University, Tokyo, Japan
4) Physics department, Kanagawa University, Yokohama, Japan
5) Department of Physics, Konan University, Kobe, 658-8501, Japan
6) ICRC, University of Tokyo, Kashiwa, Chiba, Japan
7) High energy division, Ecole Polytechnique, Palaiseau, France
8) LBNL, Berkeley, California, USA
9) INFN, Univ. di Firenze, Firenze, Italy
10) INFN, Univ. di Catania, Catania, Italy
11) IFIC, Centro Mixto CSIC-UVEG, Valencia, Spain
12) CERN, Geneva, Switzerland

This paper has been orally presented by Yasushi Muraki on behalf of the LHCf collaboration. (e-mail: muraki@konan-u.ac.jp)

Roman pot (the left panel) and by the 4" calorimeter located at outside the beam pipe (the middle panel). The right side panel of Figure 3 corresponds to the invariant mass distribution of two photons where one photon was captured by the 4" calorimeter located inside the Roman pot and the other photon was detected by the 4" calorimeter located at the outside of the beam pipe. The rapidity distribution of the neutral pions is shown in Figure 4 together with the ISR data [4] for charged pions. As shown in Figure 4, the UA7 experimental data is consistent with the Feynman scaling in the very forward region.

In closing this section we make two remarks. The UA7 experiment used a silicon strip calorimeter. Inspired by our experiment[5], A group at SLAC developed a large scale silicon calorimeter for rocket launch into space. The satellite is now named Fermi, and formerly known as GLAST. In an early GLAST report, our experimental idea was acknowledged and we are very proud of it.

The other remark is on the relationship of the density distribution of the secondary particles, $d\sigma/d\eta$ obtained by the UA5 collaboration and UA7 collaboration. The pseudo rapidity distribution $d\sigma/d\eta$ of the UA5 experiment connected smoothly with that of the UA7 experiment when the leading particles were removed from the UA5 data. This was noticed by Yoshiaki Yamamoto [6]. We present the result in Figure 5.

## 3. NEW SUBJECTS ON SUPER HIGH ENERGY PHENOMENA

Again we need to perform the same type of experiment as UA7 to measure the very forward production cross-section of neutral pions emitted at the LHC. Many new problems have arisen in the more than twenty years since UA7. One problem is again related to the composition of primary cosmic rays in the energy region of $10^{17}$-$10^{19}$eV. The other is a famous problem whether or not cosmic rays beyond the GZK cut-off exist.

Cosmic rays enter the atmosphere and produce nuclear cascade showers. The shower maximum for electromagnetic cascade showers can be written as $X_{max} \approx \ln(E_0/\varepsilon)$ where $E_0$ is the primary cosmic ray energy and $\varepsilon$ represents the critical energy in the air (84.2 MeV). However in the case of nuclear cascade showers, the shower maximum depends on the nuclear interaction model. If the shower development can be well described by the Monte Carlo simulation code SYBILL, then the primary composition of cosmic rays does not change over a wide range of cosmic ray energy from $2 \times 10^{14}$ to $3 \times 10^{19}$ eV. However if the shower development can be described by the simulation code DPMJET 2.5, the composition of cosmic rays must change drastically above $10^{16}$eV from proton dominated to heavy nucleus dominated. These results are presented in Figure 6. The conclusion then is that the composition of primary cosmic rays is completely model dependent and therefore it is very important to establish the production cross-section of neutral pions emitted in the very forward region with the Feynman variable $X_F \geq 0.1$.

Another important problem has occurred in the energy region of the highest energy cosmic rays, $E_0 \geq 10^{20}$ eV. It is expected that cosmic rays with energy higher than $4 \times 10^{19}$eV are not present in our galactic halo if the halo is filled by a magnetic field $\approx 3$ micro gauss. If the magnetic field does not exist in the galactic halo and only exists in the disk, it is even more difficult to imagine that high energy cosmic rays with energy above $4 \times 10^{19}$ eV are present. In addition we cannot expect to find extragalactic sources of ultra high energy cosmic rays >$10^{20}$eV owing to the Greisen-Zatsepin-Kuzmin cut off due to interaction with the 3K microwave background. However in spite of these expectations the AGASA air shower experiment has reported air showers with energy exceeding $10^{20}$eV which are inconsistent with the GZK cutoff. More recently the Hi Res and AUGER experiments show a declining feature of the energy spectrum at the highest region, which is consistent with the GZK cut-off. These results are presented in Figure 7. The origin of super gigantic air shower is still a matter of current physics debate.

The difference of the energy spectrum between AGASA, Hi Res and AUGER may arise from a systematic error between three different detection methods: AGASA measures the air showers with a surface array detector while Hi Res measures it by atmospheric fluorescence or the total photon detection technique. On the other hand the AUGER collaboration measures air showers with surface water tank Cherenkov detectors and atmospheric fluorescence detectors. It is said that the separation of muons from electrons is difficult with a water tank detector. The total photon detection method is independent of the nuclear interaction model in principle, but it includes other experimental biases such as the attenuation of photons in the atmosphere due to the amount of the aerosol and/or an ambiguity in determining the position of the shower center. A surface detector like AGASA has an unavoidable systematic error owing to the dependence of shower energy on a nuclear interaction model. In fact if we could shift the horizontal axis of the AGASA data toward 20% less energy, and the AUGER data toward 20% higher energy, then the three experimental results coincide with each other. To understand and resolve this problem, we need again to make a "new UA7 experiment" at the LHC in order to calibrate the nuclear interaction models at the highest energy presently possible.

## 4. THE LHCF EXPERIMENT

Two 44 radiation length compact tungsten calorimeters have been installed at zero degree collision angle +/- 140m from the interaction point of IP1. The shower calorimeters are equipped with position sensitive layers for measuring the shower axes; scintillation fiber detectors on the Arm 1 side and silicon strip detectors on the Arm 2 side. To minimize the possibility of events with multiple photons entering a calorimeter the calorimeters have been divided into two towers. The cross sectional area of the towers are 2x2 and 4x4cm$^2$ for the Arm 1 detector and 2.5x2.5 and 3.2x3.2cm$^2$ for the Arm 2 detector

At +/-140m from IP1 there is a Y shaped vacuum chamber where the beams transit from two beams in a single beam pipe facing the IP and two beams in two beam pipes facing the arcs of LHC. The LHCf calorimeters are located in the 96mm gap behind the crotch in the Y. The proton beams and the secondary charged particles are bent by the D1 beam separation dipoles located +/-20 m away from the intersection point so that only neutral particles photons, neutrons and neutral kaons reach the LHCf calorimeters.

The LHCf experiment will be carried out during low luminosity operation of the LHC $\mathcal{L}$ =5x$10^{28}$-1x$10^{30}$/$cm^2$sec in order not to incur excessive radiation damage to the LHCf detectors. According to our simulations, after an exposure of 2 minutes at $\mathcal{L}$ =$10^{29}$/$cm^2$sec, we will have enough statistics to obtain significant scientific results. To obtain a large enough sample of events with a reconstructed neutral pion mass we about 2 hour of exposure time during which we will have accumulated 40,000 such events. This is enough events to distinguish between the Monte Carlo models used for simulation of super high energy collisions.

In the summer of 2004, we performed experiments on the CERN H4 beam line. A prototype calorimeter was exposed to beams of muons, electrons and protons. These experiments measured the energy resolution of the calorimeter and the position resolution of the scintillation fiber detector. The measured energy resolution was ≈4% at 200 GeV and close to the Monte Carlo simulation. The position resolution was 200 microns. We then used this data in a Monte Carlo calculation of the neutral pion mass resolution of LHCf, obtaining $\Delta m/m$ = 4%. It was clearly demonstrated that the energy and the position resolutions of the prototype detector were sufficient for obtaining the results we need discriminate between the ultra high energy interaction models used in Monte Carlo codes. Details of these experiments have been previously published [9].

Next we discuss possible backgrounds for the LHCf experiment. The first is forward production of neutral particles by secondary particles interacting with the beam pipe. The energy spectra of these particles have been calculated and most of them found to have very low energy compared to primary neutral particles produced in collisions at the IP. The fraction of gammas from secondary particle interactions with the beam pipe having $E_\gamma$>200 GeV is 1%. The high energy neutron background is expected to be very small. Figure 8 shows a Monte Carlo prediction on the background induced by the beam pipe (calculated by K. Kasahara).

The background from beam-gas collisions has been estimated by H. Menjo in the following way. At the beginning of the LHC operation, the residual gas density is expected to be dominated by methane at a level ~ $10^{13}$ molecules/$m^3$ [10]. However after machine conditioning, the residual gas density will be reduced to $10^{11}$ molecules/$m^3$ $H_2$ equivalent. The results are shown in Figure 9. (There is a report that $10^{11}$/$m^3$ $H_2$ equivalent had already been reached on Sep. 10, 2008.) These gas densities have been estimated for highest luminosity operation of the LHC, $\mathcal{L}$ =$10^{34}$/$cm^2$sec, with 2808 bunches with $10^{11}$ protons per bunch. However in the early stages during LHCf operation, the number of bunches will be reduced by ~ 100 and the protons per bunch by ~10. Therefore the expected ratio of the beam-gas collisions to the beam-beam collisions during LHCf operation is estimated to be ~5 x$10^{-4}$. We have obtained this number integrating all beam-gas collisions along a 100m straight line.

## 5. A POSSIBLE SCENARIO IN 2009

The detectors Arm 1 and Arm 2 have been already been installed inside the TAN absorber and we are waiting to receive the "go sign" from the LHC to begin taking data as shown in Figure 10. The measurements will be made during the beam commissioning phase that is planed in the early summer of 2009. The proton beams already successfully circulated around the circumference of LHC on Sep. 10, 2008. However shortly afterward an accident occurred in the superconducting magnets between sectors 3 and 4 which prevented further commissioning before the winter shutdown. These magnets are being repaired during the shutdown and commissioning is expected to resume in the spring of 2009.

After proton bunches have been injected into the LHC orbits and before the beams have been ramped up in energy the LHCf detectors will be lowered in position to observe beam-gas collisions. Then we will attempt to locate the centers of the beams and obtain the residual gas density in the straight section of the beam pipe by selecting two photon events that reconstruct the neutral pion mass. The double arm system will be very useful also for the machine people to see "the first evidence" of beam-beam collisions. Technical details can be found elsewhere [11].

Beam conditions during the commissioning phase of LHC has been specified as follows by the machine people. At first single bunches with a few $\times 10^9$ protons will be injected into the LHC. The expected luminosity is $\mathcal{L}$ = a few $\times 10^{26}$/$cm^2$sec. Then 43 bunches with a few $\times 10^{10}$ protons each will be injected. The expected luminosity will be about $\mathcal{L}$ =$10^{30}$/$cm^2$sec. We will take the data during this time. It is expected that the LHCf detectors will incur excessive radiation damage when the luminosity increases over $\mathcal{L}$ >$10^{31}$/$cm^2$sec. The LHCf detectors would be withdrawn before reaching luminosity that would damage them. We hope that we will be able to obtain scientific results by the end of 2009 using the LHC.

## 6. SUMMARY

With the LHCf experiment important data will be obtained that will be useful for cosmic ray and high energy physics for the calibration of Monte Carlo simulation codes. In addition to data on neutral pion production we will also obtain data on neutrons and $K_0$s and the inelasticity. This paper has been presented by Y. Muraki on behalf of the LHCf collaboration.

## ACKNOWLEDGMENTS

We acknowledge Prof. Mario Calvetti, Dr. Emmanuel Tsesmelis, Dr. Michelangelo Mangano and Dr. Carsten Niebuhr for valuable and useful discussions. Thanks are also due to Prof. Lawrence Jones for valuable discussions.

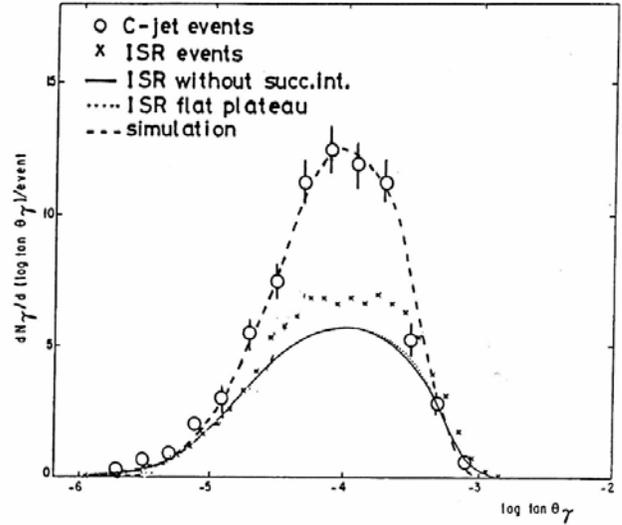

Figure 1b. The same plot of Fig. 1a, however the data points obtained by the Chacaltaya emulsion chamber slightly sifted to the left direction (o).

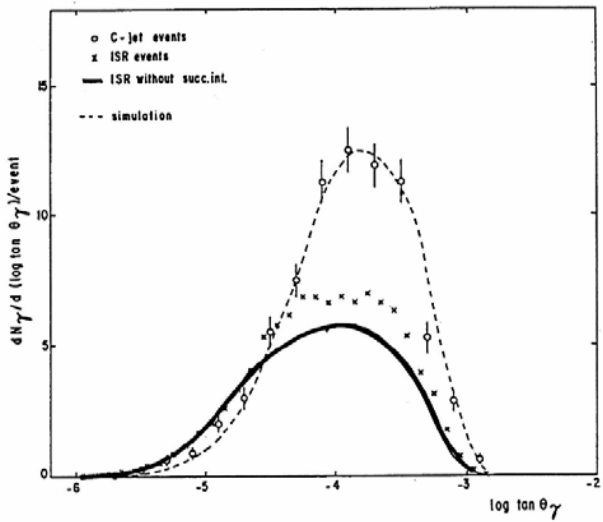

Figure 1a. The pseudo-rapidity distribution of high energy Gamma-rays observed by the Chacaltaya emulsion Chamber (o). The thick curve represents the ISR Data. Details are in the reference [3].

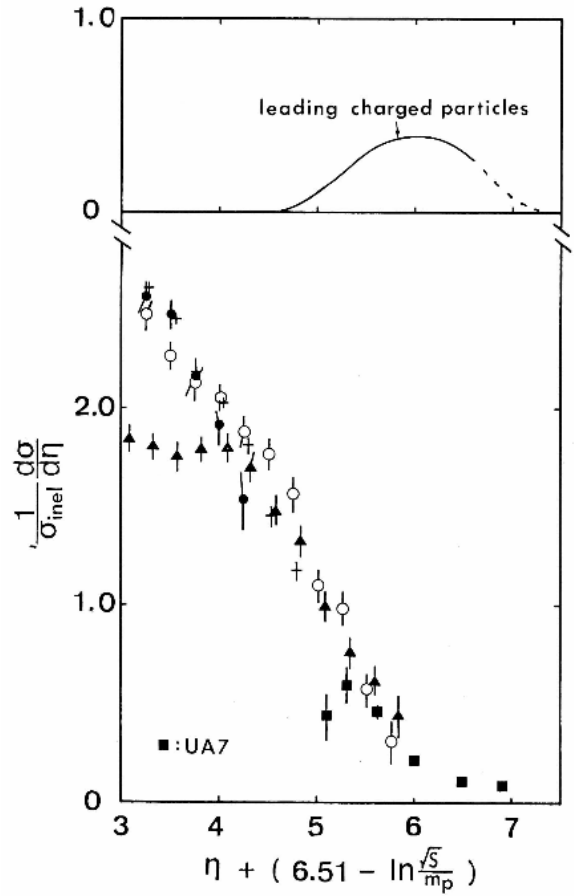

Figure 5. The UA7 data are plotted on the same plot obtained by the UA5 group. After reducing the leading particles (top panel), the pseudo-rapidity distribution can be connected smoothly. The solid squares correspond to UA7 pi zero data, while the solid triangles and white circles represent UA5 data.

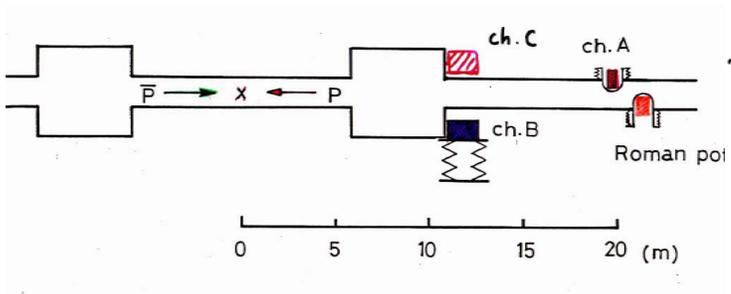

Figure 2    The configuration of the UA7 silicon calorimeter.

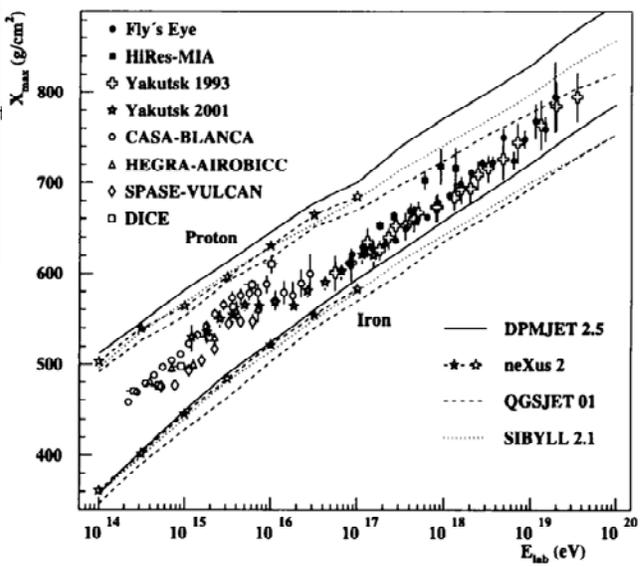

Figure 6. The shower maximum is shown as a function of the incident energy.

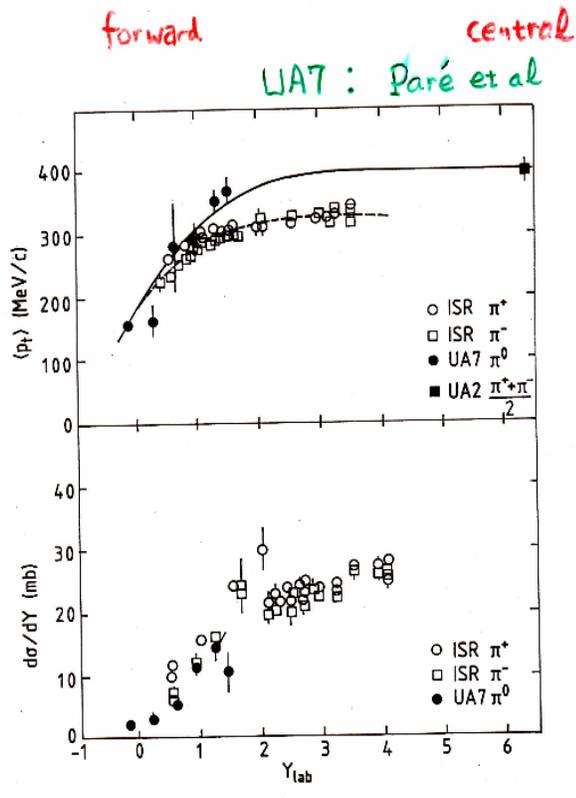

Figure 4.   The UA7 results(the right side graph).
The mean Pt distribution of the neutral pions of UA7 (top panel) and the rapidity distribution dσ /dy (bottom) are shown together with the CERN ISR charged pion data.   The scaling at very forward region is apparent.

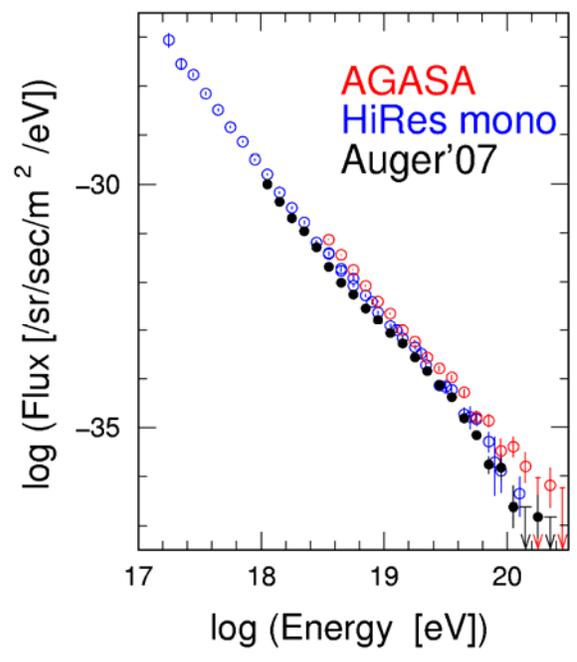

Figure 7. The energy spectrum of cosmic rays.  The red circles are the data obtained by the Akeno experiment, while blue circles are from Hi-Res.  Recent AUGER data are plotted by the black circles.

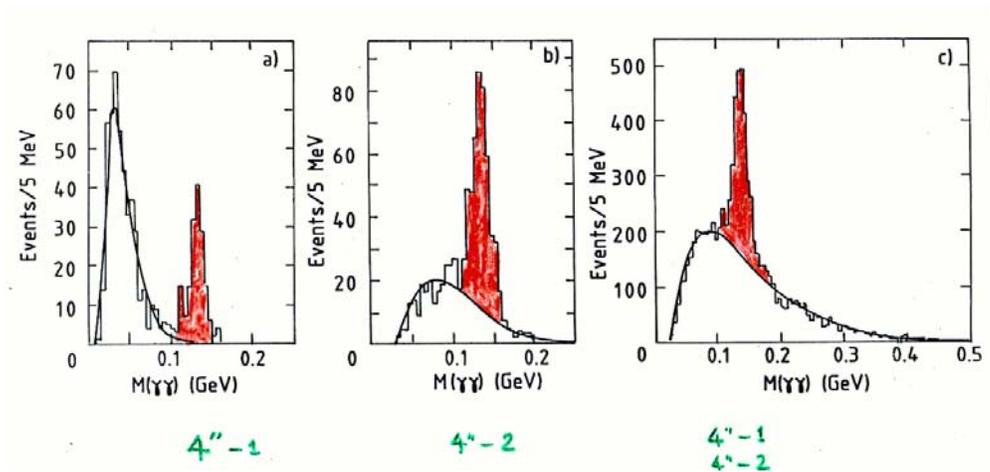

Figure 3.
The invariant mass distribution of two photons.
A clear neutral pion peak can be seen at 140 MeV.

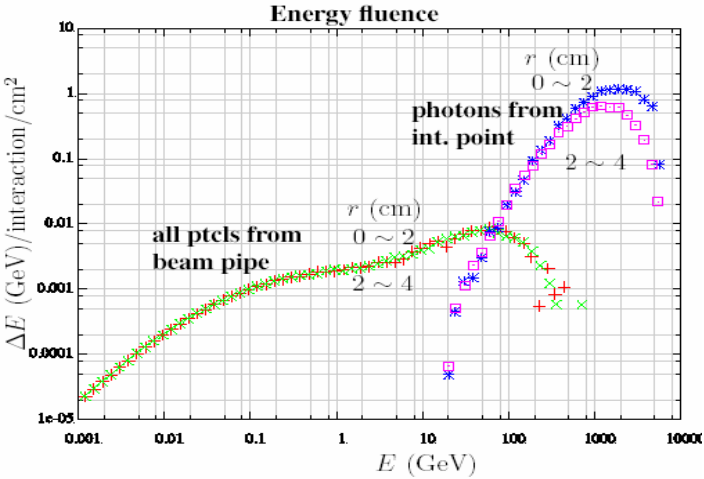

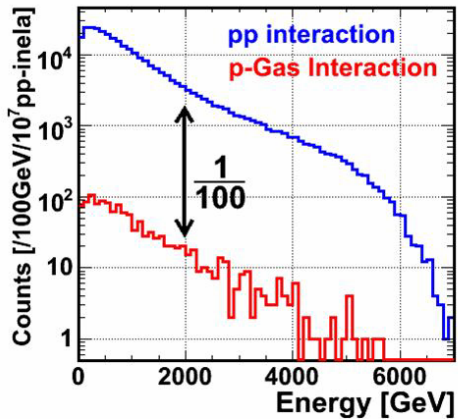

Figure 9. The estimated background by the beam-gas collisions under $4 \times 10^{12}/m^3$ H$_2$ equivalent (red line). The background is estimated to be 100 times less than the signals (the Monte Carlo calculation made by H. Menjo).

Figure 8. The energy fluency at 140 m point from the beam-beam interactions (the right side bump) and the beam pipe collisions (the left side enhancement). Vertical axis is presented as the parameter E(GeV)/interactions/cm$^2$. The vertical axis should be read as E*dN/dE*dE=EdN=ΔE.

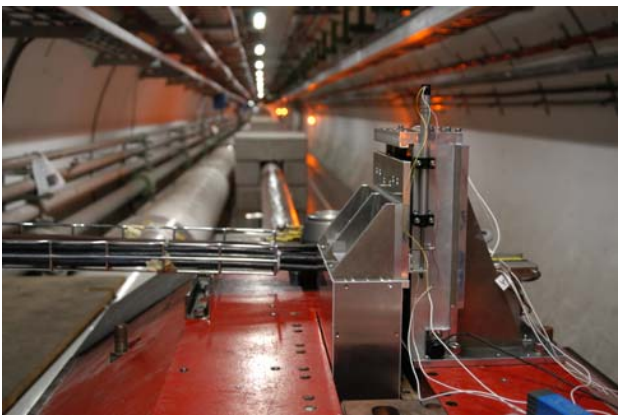

Figure 10. The LHCf detector located at the TAN, 140m away from the intersection point IP1.

.